\documentclass[%
superscriptaddress,
bibnotes,
 amsmath,amssymb,
 aps,
 prl, 
longbibliography,
eprint,twocolumn,
floatfix,
reprint
]{revtex4-2}

\usepackage{xcolor}
\definecolor{darkblue}{RGB}{0,0,150}
\definecolor{nightblue}{RGB}{0,0,100}

\usepackage{mathrsfs,dsfont,mathtools,braket,MnSymbol,graphicx}

\newcommand{\refsub}[2]{\hyperref[#1]{\ref*{#1}#2}}

\usepackage{bm}
\usepackage[
colorlinks,
citecolor=darkblue,
linkcolor=darkblue,
urlcolor=nightblue]{hyperref}

\usepackage[english]{babel}
\usepackage[babel,kerning=true,spacing=true]{microtype}

\newcommand{\be}{\begin{equation}}
\newcommand{\ee}{\end{equation}}
\newcommand{\bea}{\begin{eqnarray}}
\newcommand{\eea}{\end{eqnarray}}
\definecolor{DarkRed}{RGB}{100,0,0}
\definecolor{LightGreen}{RGB}{000,50,0}

\begin{document}
\title{
Flat-band quantum geometry from embedded quench dynamics
}

\author{Ilia Komissarov}
\email{ilia.komissarov@lsu.edu}
\affiliation{Department of Physics and Astronomy, Louisiana State University, Baton Rouge, Louisiana 70803, USA}
\author{Daniel E. Sheehy}
\email{sheehy@lsu.edu}
\affiliation{Department of Physics and Astronomy, Louisiana State University, Baton Rouge, Louisiana 70803, USA}
\author{Ahana Chakraborty}
\email{ahanachakraborty@lsu.edu}
\affiliation{Department of Physics and Astronomy, Louisiana State University, Baton Rouge, Louisiana 70803, USA}

\date{\today}

\begin{abstract}

Quantum geometry underlies a wide range of correlated phenomena in electronic systems, yet accessing the geometry of individual bands remains experimentally challenging. Advances in quantum simulators, such as ultracold atoms in optical lattices, have enabled local state preparation and site-resolved measurements, offering direct access to quantum dynamics beyond conventional transport probes. These developments naturally raise the question of whether band-resolved quantum geometry can be extracted from local occupation readouts. Here, we propose an on-site quench protocol to probe the flat-band quantum geometry via local occupation measurements in an atomic lattice. Our scheme dynamically isolates the flat-band sector by coupling a one-dimensional flat-band lattice to a two-dimensional flat-band bath. The bath selectively removes dispersive excitations, while compact localized states remain trapped due to their overlap with the flat-band modes of the bath. We show that the late-time occupation pattern directly encodes the integrated overlaps between flat-band states separated by the wavevector of the initial density pattern, thereby enabling access to integrated quantum distances between the states separated by the commensurate momenta $\delta k = 2 \pi s/r$, $s,r \in \mathbb Z$ and to the integrated flat-band quantum metric. Our results thereby establish a direct connection between quantum geometry and the long-time memory of an initial spatial pattern in flat-band systems.

\end{abstract}

\maketitle

\paragraph{Introduction.---} Quantum geometry has emerged as a unifying framework for organizing multiband effects in crystalline solids \cite{qg1, qg2, qg3, qg4, qg5, mv, Resta_1999, Mera_2021, Ozawa_2021}. It characterizes the momentum-space structure of the cell-periodic Bloch states $\ket{u(\mathbf{k})}$ and its imprint on the electronic properties of materials. Prominent examples of the quantities related to quantum geometry include the Berry curvature \cite{bc1, bc2}, which governs topological responses, and the quantum metric, which parameterizes the distance between nearby quantum states in momentum space \cite{pv} and controls quantities such as polarizability \cite{diel1, diel2, diel3} and superfluid stiffness \cite{sf1, sf2, sf3, Liang_2017, Herzog_Arbeitman_2022, Huhtinen_2022, Verma_2021}.

Quantum geometry becomes especially important when conventional measures of band dispersion, such as the Fermi velocity or inverse effective mass, vanish or are strongly suppressed. This occurs naturally in flat-band systems, where kinetic energy is quenched, and the structure of the Bloch wavefunctions becomes a dominant physical scale, with moir\'e materials providing a prototypical setting \cite{fb1, fb2, fb3, fb4}. Both transport \cite{fbt1, fbt2, fbt3, fbt4} and correlated phenomena \cite{sf1, sf2, sf3, fci1, fci2, Jackson_2015, Julku_2016, Xie_2020, Tovmasyan_2016, PhysRevB.98.134513} in these materials depend crucially on the momentum-space structure of their wavefunctions. In interacting systems, this dependence arises due to the flat-band projection that generates Hamiltonian form factors of the type $\braket{u_{\rm flat}(\mathbf{k})|u_{\rm flat}(\mathbf{k}+\delta\mathbf{k})}$. These wavefunction overlaps enter at finite momentum $\delta \mathbf{k}$ and are intrinsic to the flat band itself, whereas many existing probes of quantum geometry \cite{PhysRevB.88.064304, PhysRevB.87.245103, Ozawa_2018, Ozawa_2019}, including integrated optical-conductivity sum rules \cite{sr1, sr2, sr3}, instead access band-summed ground-state matrix elements and focus on the limit $\delta \mathbf{k} \to 0$. Developing a direct and experimentally accessible probe of the finite-momentum geometry of an individual flat band, therefore, remains an important challenge.

\begin{figure}[h!tbp]
    \centering
    \includegraphics[width=0.85\columnwidth]{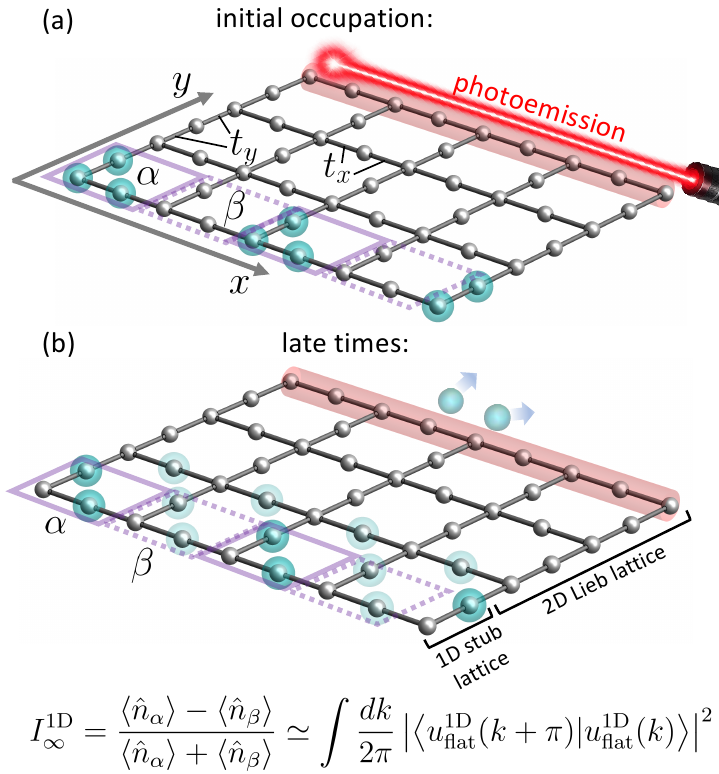}
    \caption{
    (a) We consider a two-dimensional Lieb lattice in which a single $\alpha$ bipartition of the one-dimensional subset of sites is locally populated by fermions or bosons, with a single particle per site, as indicated by the blue spheres.
    (b) At late times, dispersive particles spread to the driven edge, shown in red, where they are removed, which leads to a stationary density pattern in the vicinity of the initially populated edge. The resulting sublattice density imbalance, quantified between the $\alpha$ and $\beta$ sectors of the one-dimensional stub lattice, is related to the quantum distance between the 1D flat-band states $d^{\rm flat}_{k,k+\pi}$ \eqref{qdis}.
    }
    \label{fig1}
\end{figure}

A hallmark of flat bands is interference-induced localization \cite{par, pit, bic1, bic2, bic3, bic4, bic5, bic6, Leykam_2018, Derzhko_2015, PhysRevB.34.5208, Bergman_2008}, which we put forward as a resource for isolating and probing the intrinsic geometry of flat bands in atomic lattices \cite{exp2, exp3, exp4, exp6, Taie_2015, Jo_2012, Mukherjee_2015} through site-resolved occupation measurements. To eliminate the dispersive fermions, we use an embedding protocol in which a one-dimensional flat-band subsystem is coupled to a higher-dimensional flat-band bath (see \autoref{fig1}). 
If the bath is infinitely extended, the dispersive states escape to $y = \infty$, whereas, to prevent recurrences in realistic finite-size baths, such dispersive excitations may be removed experimentally as illustrated schematically with the laser in \autoref{fig1}.
The 2D bath, therefore, acts as a spectral filter that projects the late-time state of the 2D system onto the one-dimensional flat-band manifold (see \autoref{fig1}b). While in \autoref{fig1} we use the one-dimensional stub lattice coupled to the two-dimensional Lieb lattice, the construction is generic, as we discuss below.

The late-time state obtained after this dynamical projection retains essential information about the quantum geometry of the one-dimensional flat band. In principle, this information could be accessed through linear or nonlinear responses to applied electromagnetic fields \cite{qg1, sr1, Kolodrubetz_2017}. Here, however, we pursue a fundamentally different route that requires no transport measurement and relies only on the instantaneous site-resolved occupations, $\hat{n}_{x}$, among the most natural observables in atomic-lattice experiments \cite{n1, n2}. In particular, we consider the density imbalance between the two sublattices $\alpha$ and $\beta$ \cite{exp2, im2},
\begin{equation}
I^{\rm 1D}(t)
= \frac{\braket{\hat n_\alpha(t)}- \braket{\hat n_\beta(t)}} {\braket{\hat n_\alpha(t)}+\braket{\hat n_\beta(t)}}, \qquad \hat n_{\alpha(\beta)} = \sum_{x \in \alpha(\beta)} \hat n_x \, , 
\label{imbn}
\end{equation}
where the expectation values are computed in the instantaneous state of the system $\ket{\psi(t)}$. The bipartite density imbalance \eqref{imbn} measures the persistence of the initial density pattern over time and is related to the localization length \cite{cb}. In this work, we consider an imposed initial sublattice imbalance, where every other unit cell of the 1D lattice embedded into the 2D bath is singly occupied, with $I^{\rm 1D}(0)=1$ (see \autoref{fig1}a). We further ask: what is the long-time ($t\to \infty$) fate of this non-equilibrium quantity? In trivial flat bands (atomic limits) \cite{Nakagawa_2020, Chen_2023}, where the particles cannot propagate between the unit cells, it is natural to expect that the imbalance will not change as a function of time, staying equal to unity. However, in flat-band lattices arising from hopping interference, where the eigenstates are spread across multiple unit cells, one would expect the particles to have a finite, albeit limited, ability to propagate. As we show below, in such systems in the long-time limit, the imbalance retains a value between 0 and 1,
\begin{equation}
I^{\rm 1D}_\infty \simeq \int \frac{d k}{2\pi}\left|\braket{u^{\rm 1D}_{\rm flat}(k)|u^{\rm 1D}_{\rm flat}(k + \pi)}\right|^2 \equiv {\cal I}^{\rm 1D}\,,
\label{Iflat}
\end{equation}
determined entirely by the quantum geometry of the flat band. In the expression above, we defined the \textit{intrinsic} imbalance of a single flat band as ${\cal I}^{\rm 1D}$, which, as we show below, is equal to the integrated squared overlap of flat-band Bloch states separated by momentum $\delta k = \pi$. We note that this should be distinguished from the \textit{embedded} late-time imbalance $I^{\rm 1D}_\infty$, measured in the setup of \autoref{fig1} and possibly containing the effects induced by the 2D bath. The difference $|I^{\rm 1D}_\infty - {\cal I}^{\rm 1D}|$ can be made arbitrarily close to zero by tuning the 2D bath parameters, as we demonstrate below. The embedded imbalance thus provides a direct probe of the quantum distance between the flat-band states $\ket{u^{\rm 1D}_{\rm flat}(k)}$ and $\ket{u^{\rm 1D}_{\rm flat}(k+\delta  k)}$
\begin{equation}
d^{\rm flat}_{k,k+\delta k}=\sqrt{1 - |\braket{u^{\rm 1D}_{\rm flat}(k)|u^{\rm 1D}_{\rm flat}(k + \delta k)}|^2} \, ,
\label{qdis}
\end{equation}
via $I^{\rm 1D}_\infty \simeq 1 - \int_k (d^{\rm flat}_{k,k+\pi})^2$, where $\int_k \equiv (2 \pi)^{-1} \int dk$, with $\delta k = \pi$. We furthermore propose a series of alternative quench protocols based on the $r$-partition of the one-dimensional flat-band system, which provides access to Bloch-state overlaps at arbitrary momentum transfer $\delta k$, including the small-$\delta k$ regime governed by the quantum metric.

We note that our spectral-filtering mechanism is closely related to the physics of bound states in the continuum \cite{bic1, bic2, bic3, bic4, bic5, bic6, PhysRevA.32.3231, Miroshnichenko_2010} proposed as tunable elements for information storage \cite{bicbic1, bicbic2}. The established relation \eqref{Iflat} therefore provides an important link between quantum memory and quantum geometry --- a relation previously explored in \cite{mem1, mem2}.

The remainder of this manuscript is organized as follows. We first analyze the dynamics of a one-dimensional flat-band lattice coupled to a two-dimensional flat-band bath and demonstrate a dynamical crossover from dispersive propagation to flat-band localization driven by the decay of delocalized states. We then show how this setup isolates the flat-band imbalance and relate the bipartite imbalance to the quantum distance. Finally, we introduce multipartite imbalance observables and demonstrate how they provide access to the Brillouin zone-averaged quantum metric and to more general momentum-resolved geometric information.

\begin{figure*}[t!tbp]
    \centering
    \includegraphics[width=0.85\textwidth]{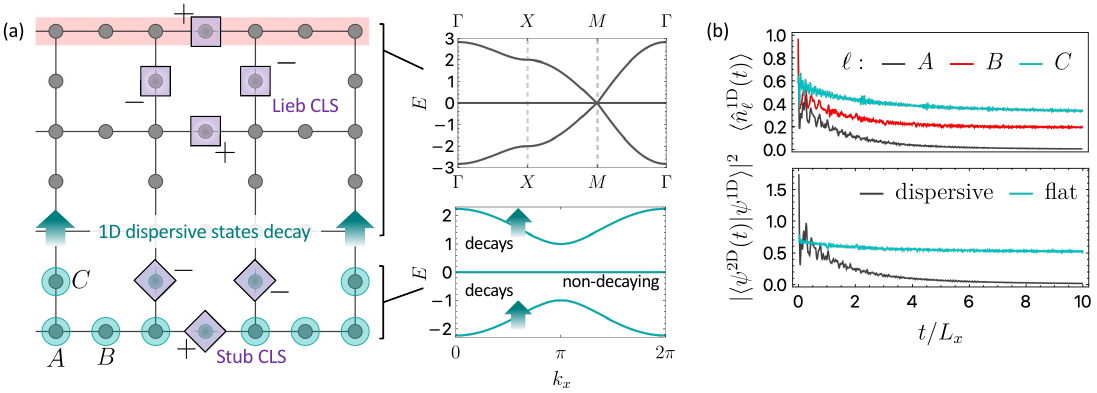}
    \caption{(a) Spectral filtering of dispersive states in a fully-filled 1D stub lattice coupled to an initially empty 2D Lieb lattice. As the system evolves, the particles, shown with blue circles, decay into the 2D, leaving a persistent localized pattern. This occurs because both the 1D system and the 2D bath host flat bands (see the band structures on the right). The subradiant nature of the 1D compact localized states (purple diamonds) originates from their finite overlap with the CLS of the 2D Lieb lattice (purple squares). (b) Time evolution of sublattice occupations $\braket{\hat n_\ell}$ \eqref{nell} and of the overlap between the 2D wavefunction $\psi$ and the 1D stub eigenstates. At late times, only $\braket{\hat n_B}$ and $\braket{\hat n_C}$ remain finite, reflecting the support of the flat-band states. Simultaneously, the overlap with dispersive 1D states decays to zero, signaling the selective depletion of dispersive modes into the 2D bath and the emergence of flat-band localization. Parameters: $L_x = 100$, $L_y = 5$, $t_x = t_y = 1$, $\gamma = 0.01$.}
    \label{fig2}
\end{figure*}

\paragraph{Flat-band bath spectral filtering.---} On-site quenches in quantum simulators typically populate multiple bands in the system \cite{dan1, dan2}. In this section, we demonstrate that the occupations of the dispersive bands can be dynamically filtered out by coupling a one-dimensional flat-band system to a two-dimensional flat-band bath.

Consider a finite segment of a 2D Lieb lattice, as shown in \autoref{fig2}a with $L_x$ unit cells in the $x$ direction and $L_y$ unit cells in the $y$ direction. The corresponding tight-binding model is given by the Hamiltonian in the $A$, $B$, $C$ sublattice basis, as labeled at the bottom of the figure,
\begin{equation}
\begin{aligned}
\hat H^{\rm 2D}
=&\,
t_x \sum_{x,y}
\hat A_{x,y}^\dagger \left(\hat B_{x,y}+\hat B_{x-1,y}\right)
 \\ +&\,
t_y\sum_{x,y}  \hat A_{x,y}^\dagger \left(\hat C_{x,y}+\hat C_{x,y-1}\right)
+
{\rm H.c.}
\, ,
\end{aligned}
\label{ham2D}
\end{equation}
where $\hat A_{x,y}$, $\hat B_{x,y}$, and $\hat C_{x,y}$ are the annihilation operators of the particles (fermions or bosons) in sublattices $A$, $B$, and $C$ in the $(x,y)$ unit cell. This model hosts a flat band at $E = 0$ generated by the compact localized states (CLS) \cite{2017PhRvB..95k5309R} with support on $B$ and $C$ sites
\begin{equation}
\ket{w^{\rm 2D}_{x,y}} = 
t_y\left(\ket{B_{x,y}}+\ket{B_{x,y+1}}\right)
-
t_x\left(\ket{C_{x,y}} + \ket{C_{x+1,y}}\right)
\, ,
\label{w2D}
\end{equation}
as shown with purple squares in \autoref{fig2}a. Here, we introduced $\hat \ell_{x,y}^{\dagger} \ket{0} = \ket{\ell_{x,y}}$ for the sublattices $\ell = A,B,C$, which should be distinguished from $\alpha$ and $\beta$, denoting the sets of alternating unit cells (see \autoref{fig1}). The 1D stub lattice, shown with blue circles in \autoref{fig2}a, connects to a 2D flat-band Lieb lattice bath. The Hamiltonian of the 1D stub lattice reads
\begin{equation}
\hat H^{\rm 1D}
=
\sum_x
\left[
t_x \hat A_x^\dagger \left(\hat B_x+\hat B_{x-1}\right)
+
t_y \hat A_x^\dagger \hat C_x
+
{\rm H.c.}
\right] \, ,
\label{ham1D}
\end{equation}
also hosting a flat band at $E = 0$ spanned by the CLS shown with purple diamonds in \autoref{fig2}a. The wavefunctions of the 1D compact localized states are
\begin{equation}
\ket{w^{\rm 1D}_{x}} = 
t_y \ket{B_{x}}
-
t_x\left(\ket{C_{x+1}} + \ket{C_{x}}\right)
\, ,
\label{w1D}
\end{equation}
Crucially, the overlap between the compact localized states of the 1D \eqref{w1D} and the 2D \eqref{w2D} lattices does not vanish in the thermodynamic limit, such that the 1D flat band states cannot completely escape into 2D and remain trapped. The existence of a non-decaying sector, as well as its absence for alternative dispersive baths, is further established through an effective Green's function analysis in the Supplemental Material \cite{SM}.

Consider a singly occupied stub-like region of the 2D Lieb lattice, as shown with blue circles in \autoref{fig2}a. The initial Fock state corresponding to this system reads
\begin{equation}
\ket{\psi^{\rm 2D}(0)} = \prod_{x} \hat A^\dagger_{x,1} \hat B^\dagger_{x,1} \hat C^\dagger_{x,1} \ket{0} \, .
\label{psi00}
\end{equation}
To prevent the recurrent dynamics expected in our finite system, we add a small absorbing term on the top row of sites, highlighted with a red rectangle in \autoref{fig2}a, defining
\begin{equation}
\hat{\widetilde H}^{\rm 2D} = \hat H^{\rm 2D} -  i \gamma \sum_{x} \left[\hat A_{x,L_y}^\dagger \hat A_{x,L_y} + \hat B_{x,L_y}^\dagger \hat B_{x,L_y} \right] \, ,
\label{htil}
\end{equation}
which simulates the laser-induced site-selective optical atom removal (see \autoref{fig1}), eliminating dispersive particles propagating to the $y=L_y$ edge.

To study the dynamics of the system in \autoref{fig2}a, we directly evolve the initial state \eqref{psi00} with the Hamiltonian \eqref{htil}, as $\ket{\psi^{\rm 2D}(t)} = e^{- i \hat {\widetilde H}_{\rm 2D} t} \ket{\psi^{\rm 2D}(0)}$ and compute the occupations within the 1D stub on all sublattices $\ell = A,B,C$
\begin{equation}
\braket{\hat n_{\ell}^{\rm 1D}}=\frac{\braket{\psi^{\rm 2D}(t)|\sum_{x} \hat \ell^\dagger_{x,1} \hat \ell_{x,1}|\psi^{\rm 2D}(t)}}{\braket{\psi^{\rm 2D}(0)|\sum_{x} \hat\ell^\dagger_{x,1} \hat\ell_{x,1}|\psi^{\rm 2D}(0)}} \, ,
\label{nell}
\end{equation}
as well as the survival probabilities within the 1D bands $|\braket{\psi^{\rm 2D}(t) | \psi^{\rm 1D}}|^2$, where the states $\ket{\psi^{\rm 1D}}$ represent the symmetrized (anti-symmetrized) product of the single-particle states diagonalizing the bosonic (fermionic) Hamiltonian \eqref{ham1D}, and have zero support outside the 1D stub \footnote{We assume the limit of small loss throughout, i.e. $\gamma$ is assumed to be negligible in comparison with other parameters in the model. In this limit, the single-particle correlation matrix description used below is justified, since the eigenstates of the Hamiltonian ${\widetilde H}_{\rm 2D}$ form an orthonormal basis, and the scattering between different modes during the time evolution can be neglected. Alternatively, one may utilize the unconditional Lindbladian particle loss description, for which \eqref{imbdef} holds exactly}. The numerical results are shown in \autoref{fig2}b, indeed revealing that the bath acts as a \textit{dynamical spectral filter}, eliminating the occupations in the dispersive bands. We note the complete decay of the states on $A$ sites, on which the CLS \eqref{w1D} have no support. As we show below, this spectral filtering mechanism provides access to the quantum geometry of the flat bands via site-resolved measurements of late-time occupations in the 1D stub.

\paragraph{Flat band density imbalance.---} The density imbalance \eqref{imbn} of a 1D system is a quantity routinely measured in quench experiments in atomic lattices \cite{exp2, im2}. Unlike transport measurements, its determination does not require an external driving field. The finite late-time value of the imbalance, $I^{\rm 1D}(t\to\infty)\equiv I^{\rm 1D}_\infty$, reveals localization in the system and can be directly related to the localization length in Anderson insulators \cite{cb}; see the End Matter for a detailed discussion. Systems with only flat-band states are localized due to lattice-induced destructive interference, resulting in a nonzero late-time imbalance, which we relate here to the integrated quantum distance between flat-band eigenstates.

\begin{figure}[t!tbp]
    \centering
    \includegraphics[width=0.85\columnwidth]{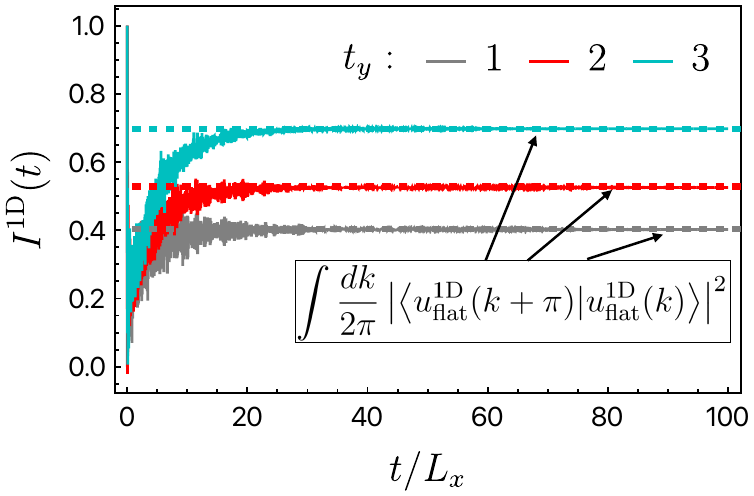}
    \caption{Embedded bipartite density imbalance for a 1D stub lattice coupled to a 2D Lieb lattice (see \autoref{fig1}) as a function of time. As dispersive states of the 1D stub are dynamically filtered out by the 2D Lieb lattice bath, in the presence of the loss term in \eqref{htil}, the imbalance saturates to a constant value. This asymptotic value closely matches the intrinsic stub imbalance, determined by the quantum distance between flat-band states shifted by $\pi$ \eqref{istub}, indicated by the dashed lines. The small residual discrepancy between the intrinsic and embedded imbalance is discussed in \cite{SM}. Parameters: $L_x = 100$, $L_y = 5$, $t_x = 1$, $\gamma = 0.01$ for all curves.}
    \label{fig3}
\end{figure}

Consider a one-dimensional system with a single flat band, which we shift to $E=0$ for convenience. We assume that every site in every other unit cell is initially singly populated: we denote the occupied sublattice by $\alpha$, while the complementary sublattice $\beta$ is initially empty. The density imbalance \eqref{imbn} between the two sublattices can be written as (see End Matter and \cite{SM})
\begin{equation}
I^{\rm 1D}(t)
= \frac{{\rm Tr}\!\left[ \hat C \, \hat U \hat P_\alpha \hat U^\dagger \right]}{{\rm Tr}\!\left[\hat U \hat P_\alpha \hat U^\dagger \right]} \, ,
\label{imbdef}
\end{equation}
where $\hat P_\alpha = \sum_{x \in \alpha, \ell} \ket{\ell_x}\bra{\ell_x}$ is the projector onto the $\alpha$ sublattice, $\hat U = e^{-i \hat H t}$ is the evolution operator in the single-particle Hilbert space $\ket{\ell_x}$, and $\hat C = \sum_{x,\ell} (-1)^x \ket{\ell_x}\bra{\ell_x} $ is the checkerboard operator, shifting momentum eigenstates by $\pi$: $\hat C \ket{k} = \ket{k+\pi}$.

We now consider an idealized time evolution protocol in which all dispersive states are dynamically removed, leaving only the unrenormalized flat-band subspace, such that 
\begin{equation}
\hat U = \hat U^\dagger = \sum_k [\ket{k}\bra{k}] \otimes [\ket{u^{\rm 1D}_{\rm flat}(k)}\bra{u^{\rm 1D}_{\rm flat}(k)}] \equiv \hat P_{\rm flat} \, .
\end{equation}
In this limit, as shown in \cite{SM}, the embedded imbalance \eqref{imbdef} reduces at late times to the intrinsic imbalance ${\cal I}^{\rm 1D}$ given by the integrated squared overlap of flat-band Bloch states separated by momentum $\pi$ [Eq. \eqref{Iflat}]. We stress that the quantity $I_\infty^{\rm 1D}$ measured in our embedding setup can differ from ${\cal I}^{\rm 1D}$ due to the renormalization of the 1D flat-band states by the 2D bath. However, we find this effect to be negligible over a broad range of parameters for the studied example of the 1D stub lattice coupled to the 2D Lieb lattice, and it can be further suppressed by tuning the system--bath coupling or by selecting a different bath, for which the flat-band state renormalization is absent, as studied in detail in \cite{SM}. The intrinsic imbalance $\cal I^{\rm 1D}$ depends solely on the parameters of the one-dimensional model; for example, for the stub Hamiltonian \eqref{ham1D}, we obtain
\begin{equation}
{\cal I}^{\rm 1D}
=
1 - \frac{4 t_x^2 |t_y|}{(2t_x^2 + t_y^2)\sqrt{4t_x^2+t_y^2}} \, .
\label{istub}
\end{equation}
In \cite{SM}, we demonstrate that this quantity is related to the localization length of the flat-band Green's function.

In \autoref{fig3}, we present the time evolution of the embedded bipartite imbalance $I^{\rm 1D}(t)$ (solid lines) and compare it to the intrinsic value \eqref{istub} (dashed lines), observing good agreement at late times. The simulations are performed for the initial configuration shown in \autoref{fig1}a, and evolved in time using \eqref{htil}.


\paragraph{Higher imbalances and the quantum metric.---} Here, we generalize the quench protocol of the previous section to quantum distances between flat-band states shifted by an arbitrary momentum $\delta k$.

\begin{figure}[t!tbp]
    \centering
    \includegraphics[width=0.8\columnwidth]{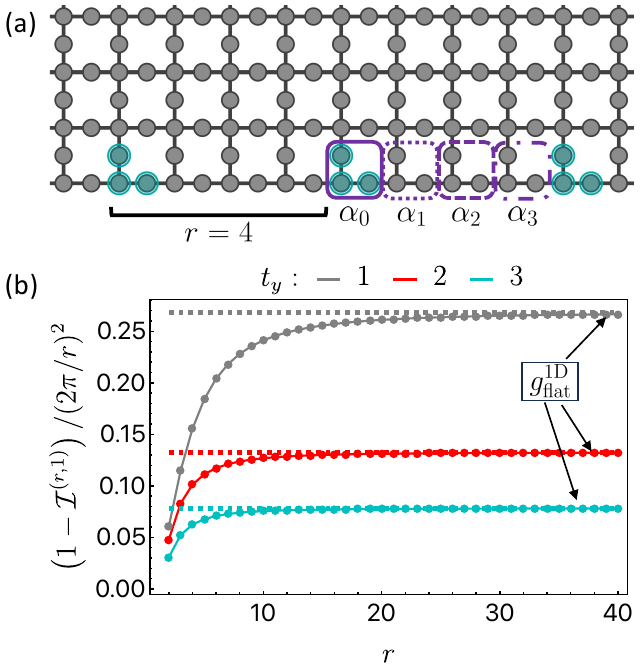}
    \caption{(a) Embedding setup used to probe the $r$-partite flat-band imbalance $I^{(r,s)}_\infty$ for a 1D stub lattice embedded into a 2D Lieb lattice: every $r$-th unit cell of the stub lattice is initially occupied (illustrated for $r=4$ with sublattices labeled as $\alpha_0$, $\alpha_1$, $\alpha_2$, $\alpha_3$). (b) The $r$-partite intrinsic imbalance for $s=1$ \eqref{irs}, subtracted from unity and normalized by the squared momentum displacement $(2\pi/r)^2$, is shown with markers (solid lines serve as visual guides). As a function of $r$, all curves converge to the flat-band quantum metric $g^{\rm 1D}_{\rm flat}$ (dashed lines), showing that the $r$-partite imbalance directly probes this quantity (see \eqref{gflatimb}). }
    \label{fig4}
\end{figure}

We define the generalized $r$-partite imbalance for an $r$-partition of a 1D lattice into $\alpha_0 \, , \ldots \, ,\alpha_{r-1}$ sublattices (see \autoref{fig4}a) as
\begin{equation}
I^{(r,s)}
=\frac{\sum_{j=0}^{r-1} \omega^{s j} \braket{\hat n_{\alpha_j}}}{\sum_j \braket{\hat n_{\alpha_j}}} \, ,
\label{Irsexp}
\end{equation}
where $\omega = e^{2\pi i/r}$, and the index $s$ runs from $1$ to $r-1$. Note that the $I^{(2,1)}$ reduces to the bipartite imbalance \eqref{imbn}. For the initial state with sublattice $\alpha_0$ singly occupied by bosons or fermions, as a result of the exact flat-band-projecting dynamics, the $r$-partite long-time imbalance becomes (see \cite{SM})
\begin{equation}
{\cal I}^{(r,s)}=\int_k \big|\braket{u_{\rm flat}^{\rm 1D}(k)\,|\,u_{\rm flat}^{\rm 1D}(k+2\pi s/r)}\big|^2 \, ,
\label{irs}
\end{equation}
which directly probes the integrated quantum distance \eqref{qdis} between flat-band states separated by momentum $\delta k = 2\pi s/r$. For our 1D stub model \eqref{ham1D}, we find
\begin{equation}
{\cal I}^{(r, s)}=1-\frac{4 t_x^2\left|t_y\right|\left(2 t_x^2+t_y^2\right) \sin ^2\!\frac{\pi s}{r}}{\sqrt{4 t_x^2+t_y^2}\left(2 t_x^4+4 t_x^2 t_y^2+t_y^4-2 t_x^4 \cos\! \frac{2 \pi s}{r}\right)}\, .
\end{equation}
In the limit of large $r$, this quantity provides access to the momentum-averaged flat-band quantum metric, which characterizes the quantum distance between infinitesimally close states,
\begin{equation}
\begin{aligned}
g_{\rm flat}^{\rm 1D}
= & 
\lim_{\delta k \to 0}
(\delta k)^{-2}\left(1 -
\int_k
\big|
\braket{
u_{\rm flat}^{\rm 1D}(k)\,|\,
u_{\rm flat}^{\rm 1D}(k+\delta k)
}
\big|^2 \right)\, \\
= & 
\lim_{r \to \infty}r^2/(2\pi)^2 (1-{\cal I}^{(r,1)}) \, ,
\end{aligned}
\label{gflatimb}
\end{equation}
where we used \eqref{irs}.
Thus, by preparing increasingly sparse occupation patterns and combining the resulting occupations with complex phase factors according to \eqref{Irsexp}, one can access the flat-band quantum metric. In the embedded system, the extraction of the intrinsic quantum metric requires control over the momentum-dependent renormalization of the surviving flat-band states, as discussed in \cite{SM}.

We illustrate the relation \eqref{gflatimb} for the 1D stub lattice in \autoref{fig4}b, where both the generalized imbalance \eqref{irs} and the flat-band quantum metric \eqref{gflatimb} are computed analytically using the 1D stub model \eqref{ham1D}. As $r$ increases, the solid curves representing the right-hand side of the equation \eqref{gflatimb} converge to the flat-band quantum metric shown with dashed lines, which assumes a simple analytical value in this model \begin{equation}
g_{\rm flat}^{\rm 1D}=\frac{t_x^2\left(2 t_x^2+t_y^2\right)}{\left|t_y\right|\left(4 t_x^2+t_y^2\right)^{3/2}} \, .
\end{equation}
Therefore, our generalized embedding protocol, after the subtraction of the bath-induced effects (see \cite{SM}), enables direct experimental access to arbitrary integrated quantum distances and to the quantum metric of flat-band states.

\paragraph{Conclusion.---} We introduced a lattice embedding protocol that dynamically projects a one-dimensional system onto its flat-band manifold. Specifically, we showed that when an occupied 1D flat-band lattice is embedded in a 2D flat-band bath, dispersive excitations hybridize with the bath and decay, while flat-band states remain confined, provided that the compact, localized states of the two systems overlap. This mechanism dynamically stabilizes the flat-band sector following a quench from an arbitrary on-site Fock state, thereby enabling the controlled study of localization and quantum geometry within an isolated flat band.

As an example of an accessible quantity, we introduced the intrinsic bipartite flat-band imbalance and demonstrated that its late-time value is directly determined by the integrated squared quantum distance between flat-band states shifted by $\delta k = \pi$. We further generalized this construction by defining $r$-partite imbalances, showing that they provide direct access to arbitrary integrated quantum distances and, in the long-wavelength limit, to the flat-band quantum metric. Our proposed approach to probing quantum geometry does not rely on transport measurements, but instead uses only on-site occupations—the natural observables in atomic-lattice experiments. This makes it possible to access quantum distances between states separated by a finite momentum $\delta k$. In contrast, conventional linear-response probes based on electromagnetic fields couple through the position operator $\hat x$ and therefore involve matrix elements of the form $\braket{u|\partial_k u}$, which characterize the quantum distance between infinitesimally close in $k$-space states.

Our findings provide a striking manifestation of quantum geometry in the dynamics of nonequilibrium systems. In general, the properties of systems driven out of equilibrium are governed by the nonequilibrium steady state, which can differ substantially from those of the ground state. Remarkably, in the quench protocol introduced here, the quantum geometry of the steady state is well approximated by that of the fully occupied static flat band. This correspondence enables direct access to quantum-geometric properties of solids in quantum simulators. Conversely, our results demonstrate that quantum geometry can have measurable consequences for nonequilibrium dynamics. In particular, compact localized states are widely used as memory elements, especially when realized as bound states in the continuum (BICs). The connection between memory retention and quantum geometry uncovered in this work may therefore provide a route toward designing improved quantum memory devices based on BICs. Specifically, we find that the quantum metric is directly related to the memory of the initial density wave pattern.

The proposed embedding protocol is readily implementable in cold-atom platforms, where two-dimensional flat-band lattices have already been experimentally realized \cite{Jo_2012, Taie_2015, PhysRevLett.118.175301, Taie_2020, PhysRevX.13.021042, Meng_2023}, and other relevant lattice implementations are discussed in \cite{SM}. Site-resolved imaging naturally provides access to the occupations dictating the embedded imbalance, an intrinsic observable sensitive to flat-band quantum geometry. Our results, therefore, establish a controlled localization probe of the quantum geometry in current-generation atomic lattice experiments.

\begin{acknowledgments}
\emph{Acknowledgments ---} DES acknowledges support from NSF award PHY-2608230. AC acknowledges support from NSF award PHY-2609899.
\end{acknowledgments}

\bibliography{lit}

\pagebreak
\newpage
\onecolumngrid
\section{End matter}
\label{em}
\twocolumngrid

\paragraph{Density imbalance.---}

\begin{figure}[t!tbp]
    \centering
    \includegraphics[width=0.7\columnwidth]{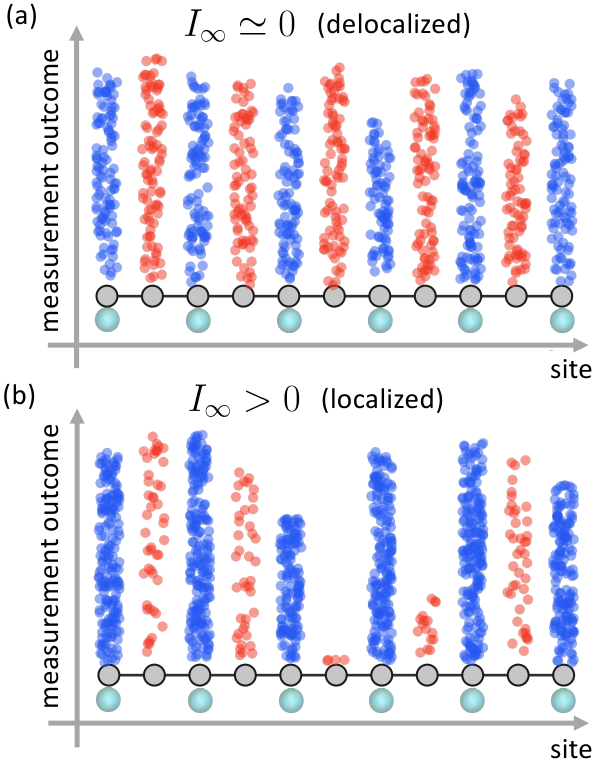}
    \caption{Behavior of imbalance in the delocalized (top) and localized (bottom) systems, with the blue spheres at the bottom indicating the initial placement of distinguishable particles in the system. The 200 measurement outcomes of the particle density at late times $t$ in the time-evolved state $\ket{\psi(t)}$ \eqref{pst} are represented with blue and red circles. (a) In the delocalized system \eqref{chain}, at long times, both sublattices are equally occupied on average, reflecting the plane-wave character of the eigenstates of the system and the vanishing of the late-time imbalance \eqref{item}, indicating the absence of retention of the initial density pattern. (b) The same initial state evolved under the Anderson Hamiltonian \eqref{And} retains memory of the initial configuration due to localization. The measured occupations remain predominantly concentrated on the initially occupied $\alpha$ sublattice, resulting in a finite positive late-time imbalance $I_\infty$. The system size is $L = 100$, with only a subset of sites shown, and the disorder strength is $W=20$. Only one disorder configuration is shown.
    }
    \label{emfig}
\end{figure}

Density imbalance measures the persistence of an initial occupation pattern in the system. Suppose a one-dimensional periodic lattice is partitioned into two sublattices --- $\alpha$ and $\beta$ --- with the former singly occupied at the initial time $t = 0$, as shown in the example of the 1D chain with blue spheres in \autoref{emfig}. The bipartite density imbalance is defined as
\begin{equation}
I^{\rm 1D}(t)=\frac{\braket{\hat n_\alpha}-\braket{\hat n_\beta}}
{\braket{\hat n_\alpha}+\braket{\hat n_\beta}},
\qquad
\hat n_{\alpha(\beta)}=\sum_{x\in \alpha(\beta)} \hat n_x \, , 
\label{item}
\end{equation}
where the average site occupations are
\begin{equation}
\braket{\hat n_x(t)}=\braket{\psi(t)|\hat c_x^\dagger \hat c_x|\psi(t)} \, .
\label{nx}
\end{equation}
We first address the bipartite imbalance in the case of a 1D hopping chain
\begin{equation}
\hat H_{\rm chain} = \sum_x \hat c^\dagger_x \hat c_{x+1} + {\rm H.c.} \, .
\label{chain}
\end{equation}
The eigenstates in this model are extended plane waves; thus, we expect that an arbitrary initial density pattern will wash out over time, resulting in a vanishing late-time imbalance \eqref{item}.

We illustrate this statement by evolving the initial density pattern 
\begin{equation}
\ket{\psi(0)} = \left[\prod_{x \in \alpha} \hat c^\dagger_x \right]\ket{0}
\label{psi0}
\end{equation}
according to
\begin{equation}
\ket{\psi(t)} = e^{-i\hat H t} \ket{\psi(0)}\, ,
\label{pst}
\end{equation}
with $\hat H = \hat H_{\rm chain}$ \eqref{chain}.

We further numerically sample the outcomes of measuring the positions of the particles in the state $\ket{\psi(t)}$ with probabilities
\begin{equation}
p_x = \frac{\braket{\psi(t)|\hat c^\dagger_x \hat c_x|\psi(t)}}{\sum_y\braket{\psi(t)|\hat c^\dagger_y \hat c_y|\psi(t)}}\,,
\end{equation}
showcasing the results in \autoref{emfig} with red and blue circles representing the positive detection outcomes. Due to the delocalized nature of the eigenstates in the model \eqref{chain}, the number of particle detections on both sublattices is roughly equal, resulting in $\braket{\hat n_{\alpha}} \simeq \braket{\hat n_{\beta}}$ at late times, and, consequently, $I^{\rm 1D}_{\infty} \simeq 0$ (see \eqref{item}). Thus, models with delocalized states yield vanishing late-time density imbalance, indicating that all memory of the initial pattern averages out.

The behavior of imbalance is qualitatively different in localized systems. Consider the one-dimensional Anderson model
\begin{equation}
\hat H_{\rm And} = \sum_x \hat c^\dagger_x \hat c_{x+1} + {\rm H.c.}
+ \sum_x \varepsilon_x \hat c^\dagger_x \hat c_x \, ,
\label{And}
\end{equation}
where the on-site disorder potential $\varepsilon_x$ is drawn independently from a uniform distribution on the interval $[-W/2,W/2]$. In this case, the spreading of the initial occupation pattern is suppressed due to the localized nature of the eigenstates. As illustrated in \autoref{emfig}b, particles in the evolved state \eqref{pst} remain predominantly detected on the initially occupied $\alpha$ sublattice, which results in a finite late-time imbalance $I^{\rm 1D}_\infty>0$. If the disorder average $\langle\!\langle \cdots \rangle\!\rangle$ is additionally performed, this imbalance can be related to the localization length $\xi$, characterizing the spatial decay of the disorder-averaged single-particle Green's function \cite{cb}, as
\begin{equation}
\langle\!\langle I^{\rm 1D}_\infty \rangle\!\rangle \simeq \tanh(\xi^{-1}) \, .
\end{equation}
Therefore, in disorder-localized systems, the imbalance directly probes localization effects induced by impurities: in the strongly localized regime $\xi \ll 1$, one finds $\langle\!\langle I^{\rm 1D}_\infty \rangle\!\rangle \simeq 1$, indicating that the initial density pattern remains frozen, while in the weakly localized regime $\xi \gg 1$, the imbalance approaches zero, corresponding to nearly uniform occupation.

The expression for the bipartite imbalance \eqref{item} greatly simplifies for non-interacting Hermitian systems, since
\begin{equation}
e^{i\hat H t}\hat c_x^\dagger \hat c_x e^{-i\hat H t}
=
e^{i\hat H t}\hat c_x^\dagger e^{-i\hat H t}
e^{i\hat H t}\hat c_x e^{-i\hat H t} \, .
\end{equation}
Using the time-evolved annihilation operator
\begin{equation}
\hat c_x(t)=e^{i\hat H t}\hat c_x e^{-i\hat H t}
=
\sum_y U_{x,y}(t)\hat c_y\, ,
\end{equation}
where $U_{x,y}(t) = \braket{x|e^{-i\hat H t}|y}$ represents the evolution operator in the single-particle Hilbert space; we rewrite the local density expectation value \eqref{nx} as
\begin{equation}
\begin{aligned}
\braket{\hat n_x}
=
\sum_{y,z}
U_{x,y}(t) & U^\dagger_{z,x}(t) \\ \times & 
\braket{
0|
\Bigl(\prod_{x' \in \alpha}\hat c_{x'}\Bigr)
\hat c_z^\dagger \hat c_y
\Bigl(\prod_{x'' \in \alpha}\hat c^\dagger_{x''}\Bigr)
|0
} \, .
\end{aligned}
\end{equation}

Defining the projector onto site $x$ as $\hat P_x = \ket{x}\bra{x}$ and the projector onto the initially occupied sublattice $\alpha$ as $\hat P_\alpha = \sum_{x \in \alpha} \ket{x}\bra{x}$, we find
\begin{equation}
\braket{\hat n_x}={\rm Tr}\left[\hat P_x\hat U(t)\hat P_\alpha\hat U^\dagger(t)\right] .
\end{equation}
Using this relation, we write the bipartite imbalance \eqref{item} in a compact form as
\begin{equation}
I^{\rm 1D}(t)
=
\frac{
{\rm Tr}
\left[
\hat C
\hat U(t)
\hat P_\alpha
\hat U^\dagger(t)
\right]}{{\rm Tr}\!\left[\hat U(t) \hat P_\alpha \hat U^\dagger(t)\right]},
\label{itc}
\end{equation}
where
\begin{equation}
\hat C=\sum_x (-1)^x \ket{x}\bra{x}
\end{equation}
is the checkerboard operator, which shifts the crystal momentum of a periodic wavefunction by $\pi$. Equation \eqref{itc} serves as the starting point for relating the bipartite imbalance to the quantum geometry explored in the main text.

\onecolumngrid
\pagebreak
\appendix
\begin{center}
\textbf{\large Supplementary information for\\ ``Flat-band quantum geometry from embedded quench dynamics"}
\\[6pt]
Ilia Komissarov, Daniel E. Sheehy, Ahana Chakraborty
\end{center}

\renewcommand{\thefigure}{\arabic{figure}}
\renewcommand{\figurename}{Supplementary Figure}
\renewcommand{\tablename}{Supplementary Table}
\setcounter{figure}{0}

This Supplementary Information is organized as follows. We first discuss the relation between the flat-band density imbalance and other quantities encoding localization, including the quantum distance, quantum metric, and localization length. We consider both bipartite and multipartite definitions of the imbalance. Next, we derive effective descriptions of different one-dimensional lattices coupled to different two-dimensional baths, arguing that filled one-dimensional systems generically radiate into the two-dimensional bath unless both subsystems host flat bands. We further show how tuning the bath parameters enables isolation of the intrinsic flat-band imbalance. Finally, we demonstrate how generic 1D flat-band lattices can be connected to form a 2D flat-band lattice, thereby allowing us to probe the flat-band-projected quantum geometry.

\section{Relation between flat band density imbalance, quantum distance, and quantum metric}
\label{appa}

In this appendix, we show that the long-time density imbalance in flat-band systems probes the quantum distance between Bloch states, sensitive to the real-space localization length of the Green's function of the system. We generalize this construction to multipartite imbalances, relating them to the quantum distance and the quantum metric of the flat band.

\subsubsection{Bipartite imbalance and quantum distance}
\label{ss1}

\begin{figure}[t!tbp]
    \centering
    \includegraphics[width=0.9\columnwidth]{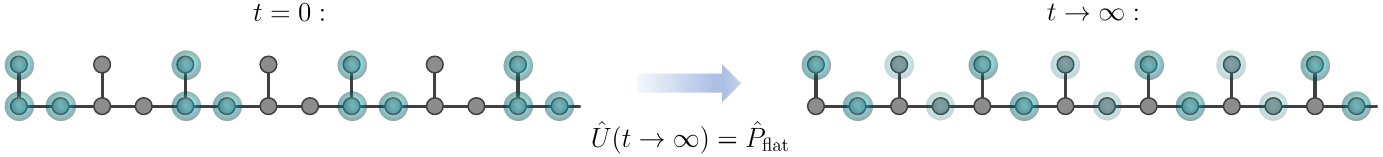}
    \caption{
    We consider the initial occupation of the flat band stub lattice with Hamiltonian \eqref{ham1Dapp}, in which every other unit cell consisting of 3 atoms is singly occupied. The time evolution $\hat U  = e^{-i \hat H t}$ is chosen in such a way that the initial density pattern at late times is projected onto the flat band manifold, which results in the bipartite imbalance related to the quantum distance between the flat band states \eqref{stubimbapp}. Note the absence of the late-time occupation of the central $A$ sites, on which the flat band states \eqref{fbs} have no support. 
    }
    \label{appfig1}
\end{figure}

Consider a bipartition of a multiband one-dimensional lattice into sublattices $\alpha$ and $\beta$, with corresponding projectors
\begin{equation}
\hat P_{\alpha} = \sum_{x \in \alpha} \ket{x}\bra{x}, 
\qquad 
\hat P_{\beta} = \sum_{x \in \beta} \ket{x}\bra{x} \, ,
\end{equation}
where $\ket{x}$ represents a singly occupied unit cell with coordinate $x$. We initialize the system with the singly-occupied unit cell bipartition $\alpha$ (see \autoref{appfig1} for the illustration) and define the imbalance
\begin{equation}
I(t)=\frac{\braket{\hat n_\alpha}-\braket{\hat n_\beta}}
{\braket{\hat n_\alpha}+\braket{\hat n_\beta}},
\qquad
\hat n_{\alpha(\beta)}=\sum_{x\in \alpha(\beta)} \hat c^\dagger_x \hat c_x \, ,
\label{itemapp}
\end{equation}
where $\hat c_x^\dagger$ and $\hat c_x$ are the creation and annihilation operators for all bosons or fermions in the unit cell $x$. If the unit cell consists of multiple sites with respective creation operators $\hat \ell_x^\dagger$, we utilize
\begin{equation}
    \hat n_{\alpha(\beta)}=\sum_{x\in \alpha(\beta),\,\ell} \hat \ell^\dagger_x \hat \ell_x \, .
\end{equation}
In the End Matter of the main text, we showed that if only the bipartition $\alpha$ is initially populated, the imbalance can be written in the projector form
\begin{equation}
I(t)= \frac{{\rm Tr}\!\left[\hat C \hat U(t) \hat P_\alpha \hat U^\dagger(t)\right]}{{\rm Tr}\!\left[\hat U(t) \hat P_\alpha \hat U^\dagger(t)\right]}\, ,
\label{Iapp}
\end{equation}
where $\hat U(t)=e^{-i\hat H t}$ is the single-particle time-evolution operator, and
\begin{equation}
\hat C=\hat P_\alpha-\hat P_\beta
=
\sum_x (-1)^x \ket{x}\bra{x}
\label{cbo}
\end{equation}
is the checkerboard operator. 

Under perfect flat-band filtering dynamics, the quantity \eqref{Iapp} represents the intrinsic property of the 1D lattice --- the intrinsic flat-band imbalance ${\cal I}(t)$. Under this assumption, at $t \to \infty$, we write
\begin{equation}
\hat U(t\to\infty)=\hat P_{\rm flat},
\end{equation}
where $\hat P_{\rm flat} = \sum_k \ket{k} \bra{k} \otimes \ket{u_{\rm flat}(k)}\bra{u_{\rm flat}(k)}$ is the projector onto the flat-band manifold spanned by Bloch waves $\ket{k} \otimes \ket{u_{\rm flat}(k)}$. Assuming a single flat band at $E=0$, the intrinsic imbalance ${\cal I} \equiv I(t\to\infty)$ takes form
\begin{equation}
{\cal I}
=
\frac{
{\rm Tr}\!\left[\hat C \hat P_{\rm flat} \hat P_\alpha \hat P_{\rm flat}\right]
}{
{\rm Tr}\!\left[\hat P_{\rm flat}\hat P_\alpha \hat P_{\rm flat}\right]
}.
\end{equation}
Using $\hat P_\alpha=(\hat I+\hat C)/2$, with $\hat I$ representing the identity operator in the unit-cell basis, we obtain
\begin{equation}
{\cal I}
=
\frac{
{\rm Tr}\!\left[\hat C \hat P_{\rm flat}\right]
+
{\rm Tr}\!\left[\hat C \hat P_{\rm flat}\hat C \hat P_{\rm flat}\right]
}{
{\rm Tr}\!\left[\hat P_{\rm flat}\right]
+
{\rm Tr}\!\left[\hat C \hat P_{\rm flat}\right]
}.
\end{equation}

The checkerboard operator shifts momentum by $\pi$,
\begin{equation}
\hat C \ket{k}
=
\sum_x (-1)^x e^{-ikx}\ket{x}
=
\ket{k+\pi},
\label{pishift}
\end{equation}
which implies ${\rm Tr}\!\left[\hat C \hat P_{\rm flat}\right]=0$. Consequently,
\begin{equation}
{\cal I}
=
\frac{
{\rm Tr}\!\left[\hat C \hat P_{\rm flat}\hat C \hat P_{\rm flat}\right]
}{
{\rm Tr}\!\left[\hat P_{\rm flat}\right]
}.
\end{equation}
Using \eqref{pishift}, we find
\begin{align}
{\rm Tr}\!\left[\hat C \hat P_{\rm flat}\hat C \hat P_{\rm flat}\right]
&=
\int\frac{dk}{2\pi}\frac{dk'}{2\pi}
\,\braket{\psi_{k'}|\hat C|\psi_k}
\braket{\psi_k|\hat C|\psi_{k'}} \\
&=
\int\frac{dk}{2\pi}
\,\left|\braket{u_{\rm flat}(k)|u_{\rm flat}(k+\pi)}\right|^2,
\end{align}
where momentum conservation fixed $k'=k+\pi$. Hence,
\begin{equation}
{\cal I}
=
\frac{1}{2\pi}
\int_{-\pi}^{\pi} dk \,
\left|\braket{u_{\rm flat}(k)|u_{\rm flat}(k+\pi)}\right|^2.
\label{Iu}
\end{equation}
The late-time bipartite imbalance of an isolated flat band is therefore determined by the average overlap between flat-band Bloch wavefunctions separated in momentum by $\pi$.

As an example, consider the one-dimensional stub lattice described by the momentum-space Hamiltonian
\begin{equation}
\hat H_{\rm 1D}(k)
=
\begin{pmatrix}
0 & t_x (1+e^{-ik}) & t_y \\
t_x (1+e^{ik}) & 0 & 0 \\
t_y & 0 & 0
\end{pmatrix} \, .
\label{ham1Dapp}
\end{equation}
The corresponding flat-band Bloch wavefunctions are
\begin{equation}
\ket{u^{\rm 1D}_{\rm flat}(k)}
=
\left[
1 + \frac{t_y^2}{2 t_x^2 (1 + \cos k)}
\right]^{-1/2}
\begin{pmatrix}
0 \\
-\dfrac{t_y}{t_x(1+e^{-ik})} \\
1
\end{pmatrix}.
\label{fbs}
\end{equation}
Substituting Eq.~\eqref{fbs} into Eq.~\eqref{Iu} yields the intrinsic flat-band imbalance of the stub lattice,
\begin{equation}
{\cal I}
=
1-\frac{4\left|t_y\right| t_x^2}
{\left(t_y^2+2 t_x^2\right)\sqrt{t_y^2+4 t_x^2}} \, .
\label{stubimbapp}
\end{equation}
This quantity is always positive and bounded from above by unity. Importantly, it depends only on the parameters of the Hamiltonian \eqref{ham1Dapp}, representing the intrinsic localization property of the flat band.

\subsubsection{Bipartite imbalance and localization length}

Disorder-averaged imbalance in 1D Anderson insulators is related to the localization length characterizing the decay of the disorder-averaged single-particle Green's function at late times \cite{cb}
\begin{equation}
\langle\!\langle I_\infty \rangle\!\rangle \simeq \tanh(\xi^{-1}) \, .
\label{imband}
\end{equation}
This result is obtained by rewriting the imbalance \eqref{itemapp} at late times as
\begin{equation} 
I_{\infty}= \frac{1}{N_\alpha} \sum_{x \in \alpha} \sum_y (-1)^y \left. \langle\!\langle \left| G(x,y) \right|^2 \rangle\!\rangle \right|_{t\to\infty} \, . 
\label{Ibrbr}
\end{equation}
Upon assuming the localized behavior $\left.\langle\!\langle |G(x,y)|^2 \rangle\!\rangle\right|_{t \to \infty} \simeq e^{-2|x-y|/\xi}$, the summation above is performed exactly, which yields \eqref{imband}.

For the intrinsic flat-band imbalance considered here, the relevant scale determining the imbalance is instead set by the decay length of the clean Green's function at the flat-band energy $E=0$, as we now demonstrate.

Consider the one-dimensional stub-lattice Hamiltonian $\hat H_{\rm 1D}(k)$ introduced in Eq.~\eqref{ham1Dapp}, and its Green's function
\begin{equation}
\hat G(k,E)
=
\left[E-\hat H_{\rm 1D}(k)\right]^{-1}.
\end{equation}
Projecting into the flat band, we find
\begin{equation}
\hat G_{\rm flat}(k,E) = E^{-1} \ket{u^{\rm 1D}_{\rm flat}(k)} \bra{u^{\rm 1D}_{\rm flat}(k)}\, .
\label{Gfk}
\end{equation}
The decay of this object in real space is determined by the pole with the smallest value of $|{\rm Im}\, k| \equiv \xi^{-1}$. The pole condition is obtained by equating the normalization denominator of the flat-band wavefunctions \eqref{fbs} to zero, which yields the flat-band localization length
\begin{equation}
4 t_x^2 \sin ^2\left(\frac{i}{2 \xi}\right)+t_y^2 =0 ~\rightarrow~ \xi=\left[ 2\,\operatorname{arsinh}\!\left(\frac{|t_y|}{2|t_x|}\right)\right]^{-1} \,  .
\end{equation}
The intrinsic imbalance of the stub lattice, Eq.~\eqref{stubimbapp}, can be expressed directly in terms of $\xi$, as
\begin{equation}
1-{\cal I}
=
2\tanh\!\left(\frac{1}{\xi}\right)
-
2\tanh\!\left(\frac{1}{2\xi}\right) \, .
\label{Itantan}
\end{equation}
This dependence differs qualitatively from the Anderson-insulator result in Eq.~\eqref{imband}, reflecting the multiband nature of the flat-band system. In the limit $\xi\to0$, one finds ${\cal I}\to1$, similarly to the Anderson insulator result \eqref{imband}. However, unlike in the Anderson insulator case, the opposite limit $\xi\to\infty$ also yields ${\cal I}\to1$. This behavior arises because $\xi\to\infty$ corresponds to $t_y\to0$, where the stub sites effectively decouple from the rest of the lattice. Particles initialized on these sites, therefore, remain trapped, preventing the imbalance from vanishing despite the divergent localization length.

We note that \eqref{Itantan} can be derived purely in real space by generalizing \eqref{Ibrbr} to our multiband system, as
\begin{equation}
    {\cal I}= \frac{1}{N_\alpha} \sum_{x \in \alpha} \sum_y (-1)^y \left. {\rm Tr} \left[ \hat G_{\rm flat}^\dagger(x,y) \hat G_{\rm flat}(x,y) \right] \right|_{t\to\infty} \, ,
\end{equation}
where, in the time domain, as follows from \eqref{Gfk},
\begin{equation}
\left. \hat G_{\rm flat}(x,y) \right|_{t \to \infty}=\delta_{x,y}\ket{C}\bra{C}+(-1)^{x-y}\begin{pmatrix}0&0&0\\0&f_{x-y} t_y^2&-t_xt_y\!\left(f_{x-y}-f_{x-y-1}\right)\\0&-t_xt_y\!\left(f_{x-y}-f_{x-y+1}\right)&- f_{x-y} t_y^2\end{pmatrix} \, .
\label{Greal}
\end{equation}
where
\begin{equation}
f_r
=
\int\frac{dk}{2\pi}\,
\frac{e^{-ikr}}
{t_y^2+4t_x^2\sin^2(k/2)}
=
\frac{e^{-|r|/\xi}}
{|t_y|\sqrt{t_y^2+4t_x^2}} .
\end{equation}
Plugging \eqref{Greal} into the real-space expression for the imbalance \eqref{Ibrbr} and carrying out the geometric sums gives
\begin{align}
{\cal I}=\frac{1-e^{-1/\xi}+2e^{-2/\xi}}{1+e^{-1/\xi}}+2\left(1-e^{-1/\xi}\right)^2\sum_{r=1}^{\infty}(-1)^r e^{-(2r-1)/\xi}=\, & \frac{1-e^{-1/\xi}+2e^{-2/\xi}}{1+e^{-1/\xi}}-\frac{2e^{-1/\xi}\left(1-e^{-1/\xi}\right)^2}{1+e^{-2/\xi}}\\=\, & 1-2\tanh\!\left(\frac{1}{\xi}\right)+2\tanh\!\left(\frac{1}{2\xi}\right) \, .
\end{align}
which reproduces Eq.~\eqref{Itantan}.

\subsubsection{Multipartite imbalance and quantum metric}

In previous sections, we showed that the bipartite imbalance is determined by the Brillouin-zone average of the squared quantum distance between flat-band Bloch states separated by a momentum shift $\delta k=\pi$ (see \eqref{Iu}). We now generalize this construction by considering a more dilute preparation in which only one out of every $r$ unit cells is initially occupied. As we demonstrate below, the resulting long-time imbalance probes the quantum distance between flat-band states separated by arbitrary commensurate momentum shifts $\delta k = 2\pi s/r$, where $s,r \in \mathbb{Z}$. We further show that in the limit $\delta k \ll 1$, such generalized imbalances provide access to the quantum metric.

To this end, consider an $r$-partition of the lattice into sublattices $\alpha_0,\ldots,\alpha_{r-1}$, with corresponding unit-cell projectors
\begin{equation}
\hat P_m
=
\sum_x \ket{rx+m}\bra{rx+m},
\qquad
m=0,\dots,r-1,
\qquad
\sum_{m=0}^{r-1}\hat P_m = \hat I.
\end{equation}
We further define
\begin{equation}
\hat Q
=
\sum_x e^{2\pi i x/r}\ket{x}\bra{x}\, , \qquad
\hat Q^s \ket{k}
=
\ket{k+{2\pi s/r}} \, .
\label{shift}
\end{equation}
For $r=2$, the operator $\hat Q$ reduces to the checkerboard operator $\hat C$ introduced in \eqref{cbo}.

The projectors $\hat P_m$ may be expressed as discrete Fourier components of $\hat Q$,
\begin{equation}
\hat P_m
=
\frac{1}{r}
\sum_{n=0}^{r-1}
e^{-2\pi i mn/r}\,
\hat Q^n.
\end{equation}
In particular, if the system is initialized with only the sublattice $\alpha_0$ occupied, the corresponding projector is
\begin{equation}
\hat P_0
=
\sum_x \ket{rx}\bra{rx}
=
\frac{1}{r}\sum_{n=0}^{r-1}\hat Q^n \, .
\label{p1}
\end{equation}

We define the generalized $r$-partite imbalance as
\begin{equation}
I^{(r,s)}(t)
=
\frac{
{\rm Tr}\!\left[
\hat Q^s
e^{-i\hat H t}
\hat P_0
e^{i\hat H t}
\right]
}{
{\rm Tr}\!\left[
e^{-i\hat H t}
\hat P_0
e^{i\hat H t}
\right]
},
\qquad
s=1,\dots,r-1.
\end{equation}
After the dispersive components are eliminated via flat-band filtering, as discussed above, the late-time imbalance becomes
\begin{equation}
{\cal I}^{(r,s)}
=
\frac{
{\rm Tr}\!\left[
\hat Q^s
\hat P_{\rm flat}
\hat P_0
\hat P_{\rm flat}
\right]
}{
{\rm Tr}\!\left[
\hat P_{\rm flat}
\hat P_0
\hat P_{\rm flat}
\right]
}.
\end{equation}

Substituting the decomposition \eqref{p1} into the numerator gives
\begin{equation}
{\rm Tr}\!\left[
\hat Q^s
\hat P_{\rm flat}
\hat P_0
\hat P_{\rm flat}
\right]
=
\frac{1}{r}
\sum_{m=0}^{r-1}
{\rm Tr}\!\left[
\hat Q^s
\hat P_{\rm flat}
\hat Q^m
\hat P_{\rm flat}
\right].
\end{equation}
Since $\hat Q^m$ shifts momentum by $2\pi m/r$ [see \eqref{shift}], the trace vanishes unless the total momentum shift is zero modulo $2\pi$. Consequently,
\begin{equation}
s+m \equiv 0 \pmod r,
\qquad
\Rightarrow
\qquad
m=r-s.
\end{equation}
Therefore, only a single contribution survives,
\begin{equation}
{\rm Tr}\!\left[
\hat Q^s
\hat P_{\rm flat}
\hat P_0
\hat P_{\rm flat}
\right]
=
\frac{1}{r}
{\rm Tr}\!\left[
\hat Q^s
\hat P_{\rm flat}
\hat Q^{r-s}
\hat P_{\rm flat}
\right].
\end{equation}
Similarly, the denominator simplifies to
\begin{equation}
{\rm Tr}\!\left[
\hat P_{\rm flat}
\hat P_0
\hat P_{\rm flat}
\right]
=
\frac{1}{r}
{\rm Tr}\,[\hat P_{\rm flat}] .
\end{equation}
Combining the two expressions yields
\begin{equation}
{\cal I}^{(r,s)}
=
\frac{
{\rm Tr}\!\left[
\hat Q^s
\hat P_{\rm flat}
\hat Q^{r-s}
\hat P_{\rm flat}
\right]
}{
{\rm Tr}\,[\hat P_{\rm flat}]
}.
\end{equation}

Expanding the flat-band projector in the Bloch basis then gives
\begin{equation}
{\cal I}^{(r,s)}
=
\frac{1}{2\pi}
\int_{-\pi}^{\pi}
dk\,
\left|
\braket{
u_{\rm flat}(k)
\big|
u_{\rm flat}\!\left(
k+{2\pi s/r}
\right)
}
\right|^2 .
\end{equation}
Equivalently,
\begin{equation}
1-{\cal I}^{(r,s)}
=
\frac{1}{2\pi}
\int_{-\pi}^{\pi}
dk\,
\left[
1-
\left|
\braket{
u_{\rm flat}(k)
\big|
u_{\rm flat}\!\left(
k+{2\pi s/r}
\right)
}
\right|^2
\right].
\end{equation}
The intrinsic multipartite imbalance, therefore, directly probes the quantum distance between flat-band states separated by momentum $2\pi s/r$.

Finally, in the limit $r\gg 1$ with fixed $s=1$, the momentum shift becomes small, and the quantum distance reduces to the quantum metric. We then obtain
\begin{equation}
g
\simeq
\frac{
1-{\cal I}^{(r,1)}
}{
(2\pi/r)^2
},
\end{equation}
showing that the multipartite imbalance provides a controlled dynamical probe of the quantum metric.

\section{Quench dynamics of 1D lattices coupled to 2D lattice baths}
\label{appb}

\begin{figure}[t!tbp]
    \centering
    \includegraphics[width=0.85\columnwidth]{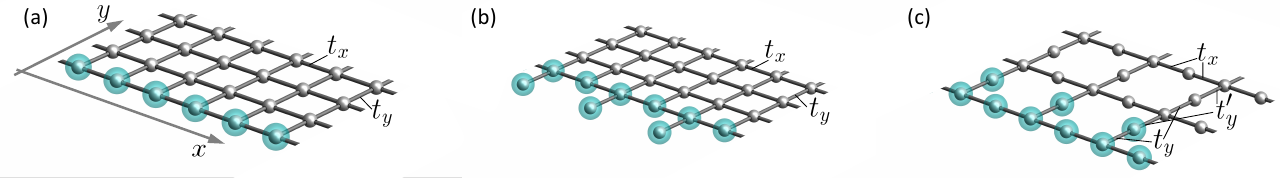}
    \caption{
    Three lattices with initial occupation shown with blue spheres: 1D chain subset occupied within a 2D square lattice (a), occupied 1D stub lattice connected to a square lattice (b), and the stub-like 1D subset of sites populated within a 2D Lieb lattice. As time passes, in configurations (a) and (b), the occupied states decay into the bulk, due to the finite overlap between the occupied on-site Fock states and the eigenstates of the infinite system. While the occupied subset in case (b) features a flat band, the occupied states are still susceptible to complete decay, since the eigenstates of the 2D bath are delocalized. In contrast, in configuration (c), a subset of the eigenstates of the 2D flat band system is localized; thus, the 1D stub lattice within the 2D Lieb lattice remains populated at $t \to \infty$.
    }
    \label{appfig2}
\end{figure}

In this appendix, we study the decay of states in different occupied 1D subsystems coupled to different empty 2D baths. Integrating out the 2D system using Schur's complement yields an effective 1D Hamiltonian that encodes the decay dynamics of the occupied 1D states. We consider 3 cases, as shown in \autoref{appfig2}: the 1D chain occupied on the boundary of the 2D square lattice bath, the occupied 1D stub at the boundary of the 2D square lattice, and finally the occupied 1D stub on the boundary of a 2D Lieb lattice bath. We further show that only in the latter case, the effective 1D Green's function hosts poles on a ${\rm Re} \, E$ axis, indicating the existence of the non-decaying states in the open quantum system.

\subsubsection{Semi-infinite 2D square lattice}
We consider a semi-infinite 2D system shown in \autoref{appfig2}a, described by the Hamiltonian
\begin{equation}
\hat H=-\sum_{x=-\infty}^{\infty}\sum_{y=1}^{\infty}t_x\left(\hat c_{x,y}^{\dagger}\hat c_{x+1,y}+\hat c_{x+1,y}^{\dagger}\hat c_{x,y}\right)-\sum_{x=-\infty}^{\infty}\sum_{y=1}^{\infty}t_y\left(\hat c_{x,y}^{\dagger}\hat c_{x,y+1}+\hat c_{x,y+1}^{\dagger}\hat c_{x,y}\right).
\end{equation}

Since the system is translationally invariant along the $x$ direction, we Fourier transform
\begin{equation}
\hat c_{x,y}=\frac{1}{\sqrt{L_x}}\sum_{k_x}e^{ik_xx}\hat c_y(k_x),
\end{equation}
which yields
\begin{equation}
\hat H=\sum_{k_x}\sum_{y=1}^{\infty}E_x(k_x)\hat c_y^\dagger(k_x)\hat c_y(k_x)-t_y\sum_{k_x}\sum_{y=1}^{\infty}\Big(\hat c_y^\dagger(k_x)\hat c_{y+1}(k_x)+{\rm H.c.}\Big)\, , \qquad E_x(k_x)=-2t_x\cos k_x.
\label{Hfourier}
\end{equation}

The equation for the Green's function of the above Hamiltonian $(E \hat I - \hat H)\hat G(E) = \hat I$ reads
\begin{equation}
\bigl[E-E_x(k_x)\bigr]G_{y,y'}(E)=\delta_{y,y'}-t_y\Big(G_{y+1,y'}(E)+G_{y-1,y'}(E)\Big).
\label{Eq:eom1}
\end{equation}

To integrate out all layers $y>1$, we decompose the Hamiltonian as $\hat H=\hat H_1 + \hat H^>+\hat V$, where
\begin{equation}
\hat V=-t_y\sum_{k_x}\left(\hat c_1^\dagger\hat c_2+\hat c_2^\dagger\hat c_1\right) \, , \qquad \hat H_1 = \sum_{k_x}E_x(k_x)\hat c^\dagger_1(k_x) \hat c_1(k_x)\, ,
\end{equation}
and $\hat H^>$ contains the remaining terms in \eqref{Hfourier}.

Denoting by $\hat G^>$ the Green's function of $\hat H^>$, we write
\begin{equation}
    \hat G =\, (E \hat I - \hat H_1 - \hat H^> - \hat V)^{-1} = \hat G^> (\hat I - (\hat H_1 + \hat V) \hat G^>)^{-1} = \hat G^> + \hat G^> (\hat H_1 +\hat V ) \hat G \, .
\label{dyson}
\end{equation}
Since the 2D bath is half-infinite and removing a single $y=1$ layer of atoms along $x$ does not change the system's Green's function. Defining
\begin{equation}
\bar E\equiv E-E_x(k_x),
\end{equation}
we find the $(1,1)$ component of Eq.~(\ref{Eq:eom1}) as
\begin{equation}
\bar E\,G_{1,1}(E)=1-t_yG_{2,1}(E),
\label{eq:preceding}
\end{equation}
where we used the boundary condition $G_{0,1}(E)=0$.
Taking the $(2,1)$ matrix element of the Dyson's equation \eqref{dyson} yields
\begin{equation}
G_{2,1}(E)=-t_y\,G_{2,2}^{>}(E)\,G_{1,1}(E).
\end{equation}
Substituting this into Eq.~(\ref{eq:preceding}) gives
\begin{equation}
G_{1,1}(E)=\frac{1}{\bar E-t_y^2G_{2,2}^{>}(E)}.
\end{equation}
Since the boundary Green's function is invariant under the removal of the $y=1$ layer of atoms, we have
\begin{equation}
G_{2,2}^{>}(E)=G_{1,1}(E),
\end{equation}
which leads to the recursion relation
\begin{equation}
t_y^2G_{1,1}^2(E)-\bar E\,G_{1,1}(E)+1=0.
\end{equation}

Solving this equation, we obtain the effective Green's function that describes the dynamics of the states in a 1D chain coupled to a 2D square lattice bath, shown with blue spheres (see \autoref{appfig2}a)
\begin{equation}
G_{\rm eff}^{\rm 1D} \equiv G_{1,1}(E)=\frac{1}{2t_y^2}\Big(\bar E-\sqrt{\bar E^2-4t_y^2}\Big).
\label{gsqfin}
\end{equation}
Fourier transforming Eq.~(\ref{gsqfin}) to the time domain yields
\begin{equation}
G_{\rm eff}^{\rm 1D}(t)=-i\Theta(t)e^{-iE_x(k_x)t}\frac{J_1(2t_yt)}{t_yt},
\end{equation}
where $J_1(x)$ is the Bessel function of the first kind, and $\Theta(t)$ is a Heaviside step function.

The survival probability of a particle initially occupying a boundary state with momentum $k_x$ is therefore
\begin{equation}
\left.\langle \hat n_{k_x}(t)\rangle\right|_{t \to \infty}=|G_{\rm eff}^{\rm 1D}(t)|^2=\left[\frac{J_1(2t_yt)}{t_yt}\right]^2\, .
\label{eq:survival}
\end{equation}
Thus, all boundary states decay algebraically with the late-time survival probability $\langle n_{k_x}(t)\rangle\propto t^{-3}$.

\subsubsection{1D stub coupled to the Semi-infinite 2D square lattice}
Here, we consider the system sketched in \autoref{appfig2}b. The stub Hamiltonian in the ${A, B, C}$ sub-lattice basis in momentum space is as follows:
\begin{equation}
\hat H^{\rm 1D} = \begin{pmatrix}
0 & t_x(1+e^{-i k_x}) & t_y \\
t_x(1+e^{i k_x}) & 0 & 0 \\
t_y & 0 & 0
\end{pmatrix} \, ,
\label{sthamapp}
\end{equation}
where $A$ is the central site, $B$ is adjacent to it along $x$ and $C$ is accessed from it by hopping along the $y$ direction. This lattice hosts a flat band at $E= 0$. Repeating the recursion trick introduced in the section above, we cast the effective 1D Green's function of the stub in the form
\begin{equation}
\hat G^{\rm 1D}_{\rm eff}(E) = (E - \hat H^{\rm 1D} - \hat V^\dagger \Sigma(E) \hat V)^{-1}\, ,
\end{equation}
where $\Sigma(E) = (\bar E-\sqrt{\bar E^2-4t_y^2})/(2t_y^2)$ is the 2D semi-infinite square bath self energy, related to \eqref{gsqfin} by a factor $t_y^2$, and $\hat V = (t_y ,~ t_y ,~ 0)$ is the vector encoding the hopping from $A$ and $B$ sites into the bath. The effective 1D Hamiltonian then enjoys energy-dependent contributions from the bath in the $A$-$B$ sector
\begin{equation}
\hat H^{\rm 1D}_{\rm eff} =  \hat H^{\rm 1D} + \hat V^\dagger \Sigma(E) \hat V = \begin{pmatrix}
t_y^2 \Sigma(E) & t_x(1+e^{-i k_x}) +t_y^2 \Sigma(E) & t_y \\
t_x(1+e^{i k_x}) + t_y^2 \Sigma(E) & t_y^2\Sigma(E) & 0 \\
t_y & 0 & 0
\end{pmatrix} \, .
\end{equation}
When the effective Hamiltonian is diagonalized, the dependence on $\Sigma(E)$, hosting a branch cut, propagates into the eigenvalues. All bands of the effective 1D systems, including the flat band, are thus also subject to the algebraic decay, similar to the case of the 1D chain on the boundary of the 2D bath (\autoref{appfig2}a).

\subsubsection{1D stub lattice connected to the 2D Lieb lattice}

\begin{figure}[t!tbp]
    \centering
    \includegraphics[width=1.\columnwidth]{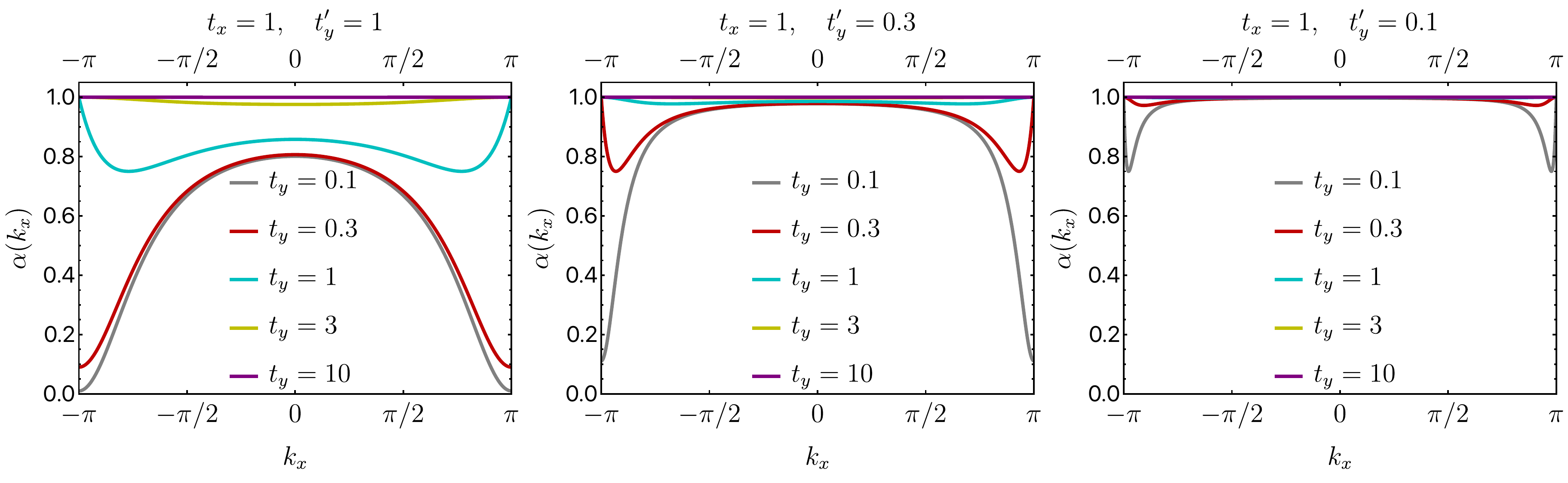}
    \caption{
    Behavior of the renormalization coefficient $\alpha(k_x)$ \eqref{akx} of the $E = 0$ 1D stub manifold connected to the 2D Lieb lattice, as sketched in \autoref{appfig2}c for three choices of the inter-stub hopping $t_y'$ with $t_x = 1$ fixed. The curves indicate that for $t_y \gg t_x$, the renormalization effect is small for any choice of the hopping $t_y'$. As $t_y'$ is decreased, individual stub lattices decouple, which results in $\alpha(k_x) \simeq 1$ for the considered $t_x$, $t_y$ parameters.
    }
    \label{appfig3}
\end{figure}

We finally proceed to the case in \autoref{appfig2}c, of a 1D stub lattice coupled to a 2D Lieb lattice by hoppings $t_y'$. We focus on the case $t_y' \leq t_y$, in which, as we show below, only the renormalized 1D flat-band sector remains occupied at late times. In the opposite case, the system \autoref{appfig2}c hosts an additional zero-mode non-decaying sector which contaminates the flat-band contribution. Below, we show that decreasing this hopping effectively decouples the occupied stub lattice from the 2D bath, thus diminishing the effect of the bath on the 1D flatband states, which brings the embedded imbalance $I$ close to the intrinsic imbalance $\cal I$ determined by quantum geometry.

 The recursion relation for the Green's function is found as outlined above
\begin{equation}
\hat{G}_{\mathrm{eff}}^{1 \mathrm{D}}(E)=\left(E-\hat{H}^{1 \mathrm{D}}-\hat{V}^{\dagger} \hat{G}_{\mathrm{eff}}^{1 \mathrm{D}} \hat{V}\right)^{-1}\, ,
\end{equation}
where the Hamiltonian $\hat{H}^{1 \mathrm{D}}$ is that of the isolated stub \eqref{sthamapp}, and the hopping matrix
\begin{equation}
\hat V =  \begin{pmatrix}
0 & 0 & t_y' \\
0 & 0 & 0 \\
0 & 0 & 0
\end{pmatrix}
\end{equation}
connects $A$ and $C$ sites of the adjacent stub and Lieb lattices. Defining $T_x=t_x(1+e^{-ik_x})$, we find the effective 1D Green's function
\begin{equation}
\hat{G}_{\mathrm{eff}}^{1 \mathrm{D}}(E)=
\frac{1}{\left(E-(t_y')^2\kappa\right)\left(E^2-|T_x|^2\right)-t_y^2 E}
\begin{pmatrix}
E\left(E-(t_y')^2\kappa\right) & T_x\left(E-(t_y')^2\kappa\right) & t_y E \\
T_x^*\left(E-(t_y')^2\kappa\right) & E\left(E-(t_y')^2\kappa\right)-t_y^2 & T_x^* t_y \\
t_y E & T_x t_y & E^2-|T_x|^2
\end{pmatrix}\, ,
\end{equation}
where $\kappa$ is found from the recursion
\begin{equation}
\kappa=\frac{E(E-(t_y^{\prime})^2 \kappa)}{(E-(t_y^{\prime})^2 \kappa)(E^2-|T_x|^2)-t_y^2 E}
\, .
\end{equation}
Remarkably, despite the coupling to the bath, the pole of the Green's function at $E = 0$  is preserved, indicating the presence of the non-decaying flat band manifold on $B$ and $C$ sublattices of the open quantum system. In the limit $E \to 0$, and for $t_y' < t_y$ we find
\be
\hat G_{\rm eff}^{\rm 1D}(E) \simeq \frac{1}{E} \frac{1}{t_y^2 + (t_y')^2 + |T_x|^2
  +\sqrt{((t_y-t_y')^2+|T_x|^2)((t_y+t_y')^2+|T_x|^2)}}
  \begin{pmatrix} 0&0&0
  \\
  0 & 2t_y^2 & -2t_y T_x^*(k_x)
  \\
  0 & -2 t_y   T_x(k_x) & 2|T_x(k_x)|^2
\end{pmatrix}
\label{Eq:green11}
\ee
This matrix is to be compared with a projector to the zero-energy manifold of the isolated stub lattice, described by the wavevectors
  \be
  \ket{u_{\rm flat}^{\rm 1D}}(k_x) = \frac{1}{\sqrt{t_y^2 + |T_x|^2}} \begin{pmatrix} 0
    \\ -t_y \\ T_x(k_x)\end{pmatrix} .
    \label{uflat1D}
  \ee
Using \eqref{Eq:green11}, \eqref{uflat1D}, we write
  \bea
 && \hat G_{\rm eff}^{\rm 1D}(E \to 0) = \alpha(k_x) E^{-1} \ket{u_{\rm flat}^{\rm 1D}(k_x)}\bra{u_{\rm flat}^{\rm 1D}(k_x)}
   \label{Eq:green11ref}
  \\
  && \alpha(k_x)=\frac{2\left(t_y^2+4 t_x^2 \cos ^2 \frac{k_x}{2}\right)}{t_y^2+\left(t_y^{\prime}\right)^2+4 t_x^2 \cos ^2 \frac{k_x}{2}+\sqrt{\left[\left(t_y-t_y^{\prime}\right)^2+4 t_x^2 \cos ^2 \frac{k_x}{2}\right]\left[\left(t_y+t_y^{\prime}\right)^2+4 t_x^2 \cos ^2 \frac{k_x}{2}\right]}},
  \label{akx}
\eea
Thus, coupling a 1D flat-band system to a 2D flat-band bath preserves the manifold of states at $E = 0$, which is retained in the open system at arbitrarily long times. However, this happens at the cost of the renormalization of the original flat-band projector of the decoupled 1D stub lattice by a momentum-dependent factor $\alpha(k_x)$. This factor makes up the difference between the intrinsic imbalance $\cal I_{\infty}^{\rm 1D}$ \eqref{Iu} and the embedded late-time bipartite imbalance, which takes the form
\begin{equation}
I_{\infty}^{\rm 1D} =
\frac{
\int \frac{dk_x}{2 \pi} \, \alpha(k_x) \alpha(k_x+\pi)
|\braket{u_{\rm flat}^{\rm 1D}(k_x)|u_{\rm flat}^{\rm 1D}(k_x+\pi)}|^2}{
\int \frac{dk_x}{2 \pi} \, \alpha^2(k_x)}\, .
\end{equation}
Nevertheless, we find that in a considerable parameter range, the effect of renormalization encoded in $\alpha(k_x)$ can be neglected. For reference, the parameters considered in \autoref{fig3} yield for the semi-infinite bath ($I_{\infty}$), as compared to the intrinsic values (${\cal I}$)
\begin{equation}
\begin{gathered}
t_x = 1 \, ,~ t_y = t_y'  = 1: \qquad I_{\infty} \simeq 0.395 \qquad {\cal I} \simeq 0.404 \, , \\
t_x = 1 \, ,~ t_y = t_y'  = 2: \qquad I_{\infty} \simeq 0.524 \qquad {\cal I} \simeq 0.529 \, ,\\
t_x = 1 \, ,~ t_y = t_y'  = 3: \qquad I_{\infty} \simeq 0.690 \qquad {\cal I} \simeq 0.697 \, ,
\end{gathered}
\label{II}
\end{equation}
which implies that our embedding scheme provides a faithful approximation of the intrinsic imbalance even without bath engineering. More generally, we showcase the behavior of the renormalization coefficient $\alpha(k_x)$ \eqref{akx} in different parameter ranges in \autoref{appfig3}. As the system-bath coupling $t_y'$ is decreased, the effect of the 2D bath on the 1D flat band states becomes negligible, and we find $\alpha(k_x) \simeq 1$. Thus, tuning $t_y' \ll t_y$ can be used to further improve on the results in \eqref{II}, achieving $I_{\infty} \to {\cal I}$ in the limit $t_y' \to 0$. Finally, one can replace the Lieb bath with a different flat-band bath, which has $\alpha(k_x) = 1$, as illustrated below.
  
\section{Constructing 2D flat band lattices from 1D flat band segments}
\label{appc}

\begin{figure}[t!tbp]
    \centering
    \includegraphics[width=0.7\columnwidth]{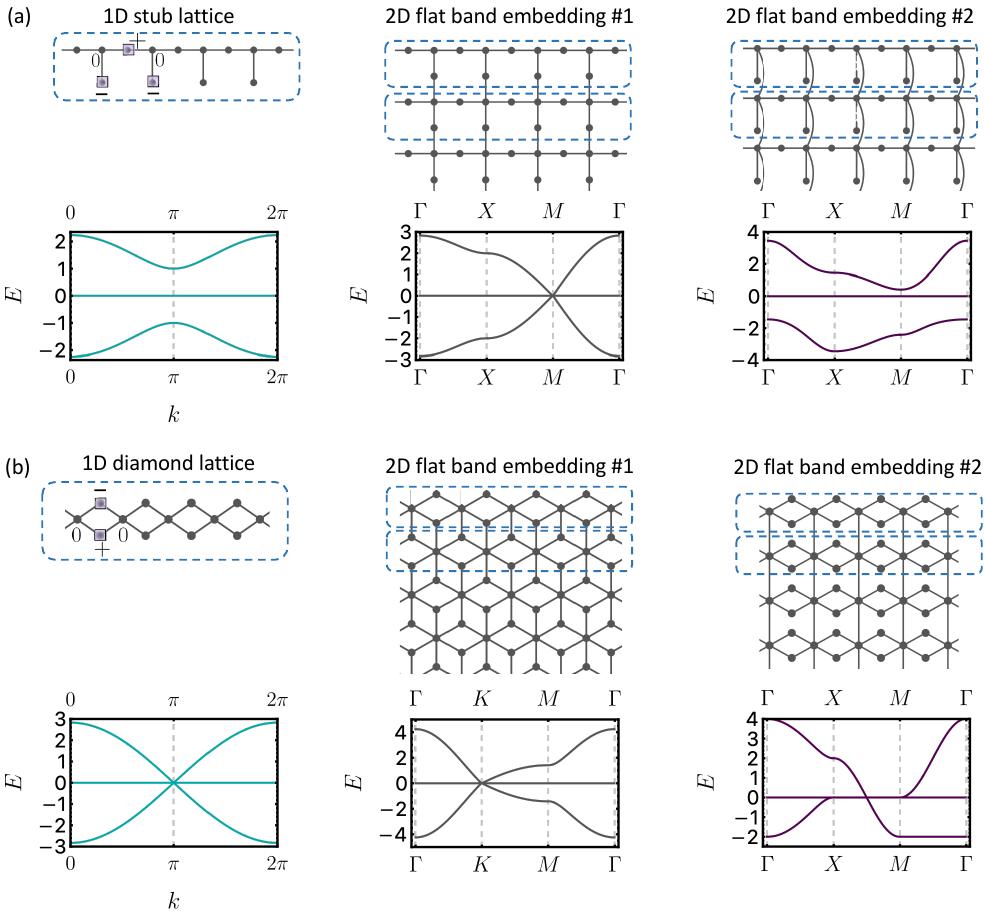}
    \caption{
    Constructing a 2D flat-band lattice using 1D flat-band segments highlighted with dashed rectangular profiles. (a) The 1D stub lattice with the compact localized state highlighted with purple markers. The sign of the compact localized state (CLS) wavefunction is indicated near each marker: the CLS avoids the central sites marked with 0. On the right, we show two types of connections between the 1D stub lattices that yield 2D flat-band lattices. Case \#1 (the 2D Lieb lattice) is obtained by connecting ``$+$'' sites of one stub lattice to ``$0$'' sites of another stub lattice with additional hoppings. Case \#2 lattice (a 2D square lattice decorated with stubs) is obtained by gluing the ``$0$'' sites of the original stub lattices. On the bottom, we sketch the dispersions of all bands of all three lattices. (b) The same procedure is illustrated for the 1D diamond lattice, resulting in the 2D dice lattice (case \#1) and another 2D flat band. In all band structure plots, all hopping elements are set to unity.
    }
    \label{appfig4}
\end{figure}

In the main text, we showed that the 1D stub lattice can be embedded into a 2D Lieb lattice, with both lattices sharing a flat band. This feature allows dynamic isolation of 1D flat-band states, providing access to the 1D flat-band quantum geometry. In this section, we argue that this embedding is not unique and that, starting from a 1D flat-band lattice, one can build a flat-band 2D lattice with the original 1D system featured as a subset (see also \cite{Morales_Inostroza_2016}).

To proceed, one should identify the compact, localized states of the 1D flat-band lattice. In \autoref{appfig4}a, one compact localized state is shown with purple markers for a 1D stub lattice, with the signs of the wavefunction shown. The central sites are avoided by the CLS, as indicated by ``$0$''. There are at least two ways to produce a 2D flat band lattice by gluing together 1D stub lattices, as sketched to the right. First, one can glue the ``$+$'' or ``$-$'' sites of one stub to the ``$0$'' sites of the other stub, and then proceed recursively. The Hamiltonian of the resulting lattice has the same structure as the original, leading to the existence of 2D compact localized states and the formation of a 2D flat band. In momentum space, we write
\begin{equation}
\hat H^{\rm 1D} = \begin{pmatrix}
0 & t_x(1+e^{-i k_x}) & t_y \\
t_x(1+e^{i k_x}) & 0 & 0 \\
t_y & 0 & 0
\end{pmatrix} \rightarrow \hat H^{\rm 2D}_{\rm Lieb} = \begin{pmatrix}
0 & t_x(1+e^{-i k_x}) & t_y(1+e^{-i k_y}) \\
t_x(1+e^{i k_x}) & 0 & 0 \\
t_y(1+e^{i k_y}) & 0 & 0
\end{pmatrix} \, .
\end{equation}
The Hamiltonian $\hat H^{\rm 2D}_{\rm Lieb}$ features a flat band, although the flat band wavefunctions are distinct from those of the 1D Hamiltonian, which results in the finite renormalization of the wavefunctions indicated by $\alpha(k_x) \neq 1$, as obtained in the previous section.

The second way to glue the 1D flat band lattices into a 2D flat band is by connecting the ``0'' sites of the $1D$ lattices, as shown on the right in \autoref{appfig4}a. This results in a square lattice with a single stub inside each plaquette
\begin{equation}
\hat H^{\rm 1D} = \begin{pmatrix}
0 & t_x(1+e^{-i k_x}) & t_y \\
t_x(1+e^{i k_x}) & 0 & 0 \\
t_y & 0 & 0
\end{pmatrix} \rightarrow \hat H^{\rm 2D}_{\rm square+stub} = \begin{pmatrix}
2 t_y \cos k_y & t_x(1+e^{-i k_x}) & t_y\\
t_x(1+e^{i k_x}) & 0 & 0 \\
t_y & 0 & 0
\end{pmatrix} \, .
\end{equation}
Interestingly, this type of coupling preserves the original 1D flat-band CLS, i.e., $\alpha(k) = 1$, and therefore is more desirable for the purpose of probing the quantum geometry of the 1D flat band. Lastly, we note that a similar gluing argument can be applied to other lattices; for example, in \autoref{appfig4}b, we consider a 1D diamond lattice, which yields two 2D flat-band lattices from its repetitive stacking and connection.

\end{document}